\documentclass[10pt,letterpaper]{article}
\usepackage{opex3}
\usepackage{color}

\renewcommand\thesubsection{\thesection.\Alph{subsection}}

\usepackage{graphicx}
\usepackage{amsmath}
\usepackage{color}
\usepackage{stfloats}
\usepackage{lipsum}

\begin{document}
\title{Microcavity with saturable nonlinearity under simultaneous resonant and nonresonant pumping: multistability, Hopf bifurcations and chaotic behaviour.  }
\author{Ivan Iorsh$^{1*}$, Alexander Alodjants$^{1}$, Ivan A. Shelykh$^{1,2,3}$}
\address{$^1$National Research University of Information Technologies,
Mechanics and Optics (ITMO), St.~Petersburg 197101, Russia}
\address {$^2$Science Institute,
University of Iceland, Dunhagi-3, IS-107 Reykjavik, Iceland,}
\address {$^3$Division of Physics and Applied Physics, Nanyang
Technological University 637371, Singapore.}
\email{$^*$i.iorsh@phoi.ifmo.ru}

\begin{abstract}
We studied optical response of microcavity non-equilibrium  exciton-polariton Bose-Einstein  condensate with saturable nonlinearity under simultaneous resonant and non-resonant pumping. We demonstrated the emergence of multistabile behavior due to the satutration of the excitonic absorbtion. Stable periodic Rabi- type oscillations of the excitonic and photonic condensate components in the regime of the stationary pump and their transition to the chaotic dynamics  through the cascade of Hopf bifurcations  by tuning of the electrical pump are revealed.
\end{abstract}
\ocis{(190.1450) Bistability, (190.5970) Semiconductor nonlinear optics }

\section{Introduction}
Exciton-polaritons are quasiparticles emerging due to the strong coupling between quantum well excitons and cavity photons in high quality factor cavities~\cite{Microcavities}. Due to their hybrid light-matter nature, polaritons exhibit set of intriguing collective effects such as high temperature Bose-Einstein condensation (BEC)~\cite{BEC0}, superfluidity~\cite{SuperFluid} and macroscopic self trapping \cite{SelfTrapp}. Those phenomena originate from very small effective mass of polaritons inherited from the photonic component combined with strong polariton- polariton interactions provided by the excitonic component. 

Together with the intriguing fundamental properties, exciton-polaritons are attractive for their perspective applications, mainly in the field of low-threshold bosonic lasers~\cite{Pol_Laser0} and all-optical integrated circuits~\cite{Pol_Circuits,Pol_Circuits2,Pol_Transistor,
Log_gate1,Log_gate2,Log_gate3,Log_gate4,Log_gate5}. The advantage of the polaritonic platform over the conventional nonlinear optical materials for the purposes of the development of optical logic elements is their ultra-strong and ultra-fast nonlinear response ~\cite{NLresp1,NLresp2,NLresp3}.

One of the important  features  of non-equilibrium exciton-polariton BEC  that can be used in practice  relays to so-called permanent Rabi oscillations  occurring  between lower and upper branch polaritons~\cite{Permanent1}  or between photonic and excitonic components of the condensate~\cite{Permanent2,Permanent3}. An enhansement of the coherence time of polariton Rabi oscillations  has been reported experimentally in Ref.~\cite{Perm_exp}.  The oscillation pattern strongly depends on the mechanism of the pumping, but the role  of different contributions to the nonlinear excitonic responce and different types of the pumping in polariton Rabi oscillations in microcavities is still not fully clarified.

In this paper we examine the problem of the onset of permanent Rabi oscillations between excitonic and photonic components in the microcavity in the presence of strong nonlinearities caused by both exciton-exciton interaction and saturation of the excitonic absorbtion provided by the effects of Pauli blocking~\cite{exc-exc1,exc-exc2,sat1,sat2}. The latter plays major role in the regime of very strong pumps. Indeed, as exciton density $n_X$ approaches the critical concentration of $1/a_B^2$, where $a_B$ is the excitonic Bohr radius, the system undergoes the Mott transition, and the excitons dissolve forming the electron-hole plasma.  We also take into account the possibility of having two types of the pump: resonant pump of the cavity mode by external laser with well defined frequency and non- resonant pump of the excitonic component provided e.g. by electrical injection of the electrons and holes into active region \cite{Pol_Laser0,Sci_Rep_PolLas}
\begin{figure}[!h]
\centerline{\includegraphics[width =
0.4\columnwidth]{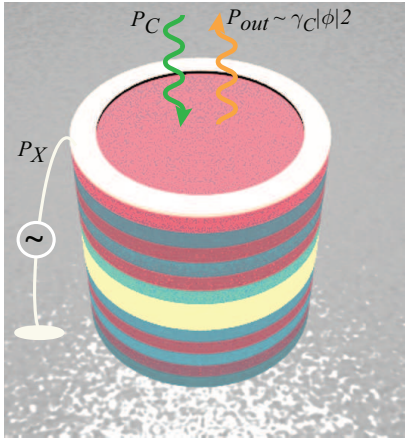}} \caption{Sketch of the system under consideration: a polaritonic microcavity is pumped simultaneously with an electrical pump $P_X$ and optical pump $P_C$. } \label{fig_1}
\end{figure}

It is well known that one of the direct consequences of the polariton nonlinear response is the optical bistability ~\cite{Bistable_pred,BS_exp_opt}. Importantly, bistable behavior was observed for the cases of both resonant optical and non- resonant pump when excitonic subsystem is pumped via the electron- hole reservoir. The mechanisms of the bistability, however, are very different in these two cases. For resonant optical pump it appears due to the resonance between interaction- induced blueshift and detuning between the energies of the cavity mode and the pumping laser. On the contrary, for the non- resonant pump bistabile behavior was reported for the electrically pumped polaritonic diode as resulting from the transition to the weak coupling regime \cite{BS_exp_elec2} or dependence of the electron-hole tunneling lifetime on the carrier density provided by the effects of the screening ~\cite{BS_exp_elec}.  

In this work we consider the coupled exciton-photon system in a microcavity configuration shown in Fig.~\ref{fig_1} subject to the \textit{simultaneous} electrical and optical pump and account for the exciton-photon coupling saturation due to the Mott transition. We reveal that the system may exhibit the multistable behaviour as well as undergo the transitions from stationary solutions to stable periodic oscillations of excitonic and photonic components and then chaotic behaviour if the intensities of the pumps are tuned. The multistable behaviour has been previously observed in the microcavities~\cite{Multi_1,Multi_2,Multi_3}, however conventionally the observation of multistability requires an account for spin(polarization) degree of freedom.  Contrary to this case, in this paper multistability emerges due to the interplay of the exciton-exciton interaction and the saturation of the exciton-photon coupling. The chaotic behaviour of exciton polaritons has been previously studied in the optically pumped polariton Josephson junctions~\cite{chaos1} and planar microcavities with briken polarization symmetry~\cite{chaos2,chaos3}. In this work the transition to the chaotic behaviour can be achieved by tailoring the electrical pump which could be advantageous for the applications in chaos communication devices~\cite{chaos_comm}.

The remaining paper is organized as follows. In section~\ref{sec1} we introduce the Hamiltonian of the system and obtain the equations of motion for the excitonic and photonic amplitudes as functions of time. In section~\ref{sec2} the analysis of the stationary solutions is performed. Section~\ref{sec3} contains the stability analysis of the stationary solutions. Moreover, the regimes of limiting cycles characterized by permanent oscillations for excitonic and photonic field amplitudes as well as Hopf bifurcations and transition to chaos through period- doubling cascade are discussed. The conclusions are presented in  section~\ref{sec4}.

\section{Model}\label{sec1}
We consider the coupled exciton-photon semiclassical Hamiltonian written as
\begin{align}
&\mathcal{H}=\hbar\omega_{C}\phi^*\phi+\hbar\omega_{X}\chi^*\chi+\frac{g}{2}|\chi|^4+\hbar\Omega_R\left(1-\lambda a_B^2|\chi|^2\right)\left(\phi^*\chi+\chi^*\phi\right),
\end{align}
where $\phi,\chi$ are the complex amplitudes of the photonic and excitonic macroscopic wave functions, respectively, $\omega_C,\omega_X$ are the frequencies of the excitonic and photonic mode, $g$ defines the strength of the exciton-exciton interaction and can be approximated by~\cite{Yamamoto} $g\approx 6E_b a_B^2$, where $E_b$ is the exciton binding energy and $a_B$ is the exciton Bohr radius. The  dimensionless constant $\lambda$ defines the efficiency of the saturation of the excitonic absorption. If $\lambda=0$ the saturation is absent and one regains the conventional Hamiltonian of coupled excitons and photons with only source of nonlinearity provided by exciton- exciton interactions which was extensively used for description of the polariton dynamics and leads to the standard bistable behavior in the regime of the resonant pump (see e.g. \cite{CiutiReview} and references therein). Our main goal will be to analyze the additional source of nonlinearity appearing for $\lambda\neq0$. Note, that Mott transition from excitons to electron-hole plasma occurs roughly when the distance between the individual excitons become comparable to their Bohr radius, and thus realistically $\lambda\approx1$.

Throughout the paper we neglect spatial degree of freedom assuming the condensate to be at zero momentum state.  Equations of motions for the excitonic and photonic amplitudes are derived from the Hamiltonian as $i\hbar\dot{\xi}=\partial{\mathcal{H}}/\partial {\xi^*}, \quad \xi=\{\phi,\chi\}$. We also will phenomenologically add the terms corresponding to the exciton and photon damping $\gamma_X,\gamma_C$, as well optical and electrical pumping terms, $P_Ce^{-i\omega t},\quad P_X\chi(1-\lambda a_B^2|\chi|^2)$ , respectively. Note, that saturation of the incoherent pump originating from the same effect of Mott transition is introduced. Making the substitution $\xi=\tilde{\xi}e^{-i\omega t}a_B^{-1}$, $\xi=\{\phi,\chi\}$ and introducing the dimensionless time $\tau=\Omega_R t$ brings us to a couple of dimensionless differential equations:
\begin{align}
&i\dot{\tilde{\phi}}=(\tilde{\delta}_C-i\tilde{\gamma}_C)\tilde{\phi}+(1-\lambda|\tilde{\chi}|^2)\tilde{\chi}+\tilde{P}_C,\nonumber\\
&i\dot{\tilde{\chi}}=(\tilde{\delta}_X-i\tilde{\gamma}_X)\tilde{\chi}+(1-\lambda|\tilde{\chi}|^2)\tilde{\phi}-2\lambda\tilde{\chi}\mathrm{Re}\left(\tilde{\phi}^*\tilde{\chi}\right)+\tilde{g}|\tilde{\chi}|^2\tilde{\chi}+i\tilde{P}_X\tilde{\chi}(1-\lambda|\tilde{\chi}|^2),
\end{align}
where $\tilde{\delta}_{C(X)}=(\omega_{C(X)}-\omega)/\Omega_R$ are detunings of the excitonic and photonic modes with respect to the laser frequency and  all the parameters with the dimension of energy were normalized to $\Omega_R$. Thereafter we remove tilde sign from the symbols to spare the notations.
\begin{figure}[!h]
\centerline{\includegraphics[width =
0.8\columnwidth]{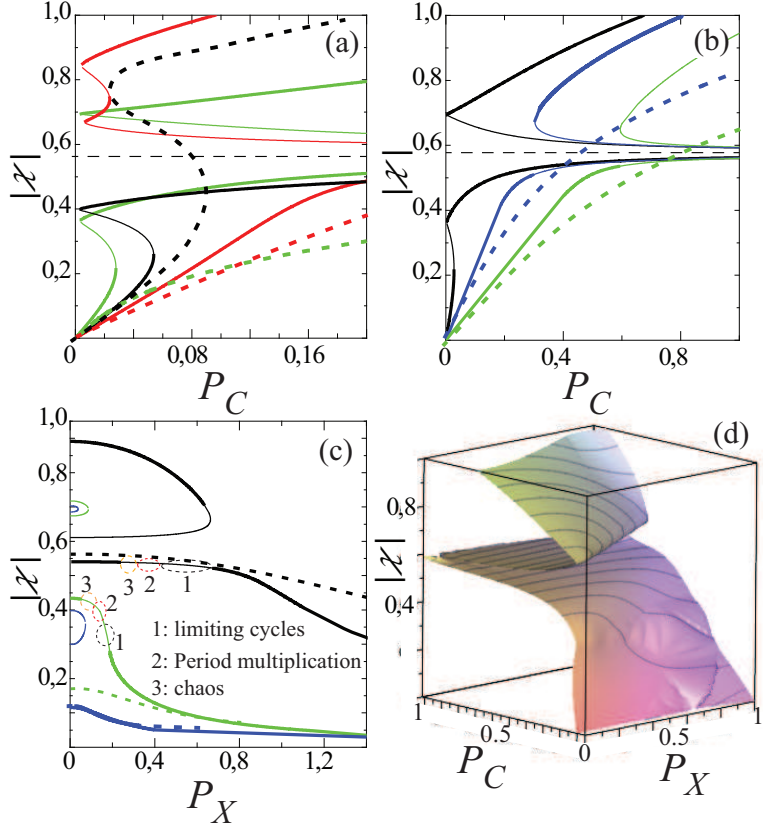}} \caption{(a) Exciton amplitude vs the optical pump power for the absent electrical pumping at the pumping detuning equal to -9 meV (green line), -2 meV (red line) and 5 meV (black line); all other parameters can be found in the text. (b) Exciton amplitude vs the optical pump power at $\delta_C=\delta_X=-9$meV at ${P}_X=0$ (black line), 1.12 (blue line) and 1.2 (green line. (c) Exciton amplitude vs the electrical pump power at $\delta_C=\delta_X=-9$meV at ${P}_C=0.2$ (green line), 0.5 (black line), and 0.02 (blue line). In figures (a-c) solid lines correspond to the saturable case $\lambda=1$ and dotted to $\lambda=0$. (d) Three dimensional plot of exciton amplitude vs electrical and optical pump for $\lambda=1$ and $ \delta_C=\delta_X=-9$~meV. }  \label{fig_2}
\end{figure}

In the stationary case $\dot{{\phi}}=0,\dot{{\chi}}=0$, the photon amplitude can be expressed via the exciton amplitude as
\begin{align}
{\phi}=-\frac{e^{i\varphi_0}}{\sqrt{{\delta}_C^2+{\gamma}_C^2}}\left(\left[1-\lambda|{\chi}|^2\right]{\chi}+{P}_C\right), \label{phiexp}
\end{align}
where $\tan\varphi_0=\gamma_C/\delta_C$. Equation~\eqref{phiexp} can then be substituted to the equation for $\chi$. We then can look for the solution for ${\chi}$ in the form ${\chi}=Xe^{i\varphi}$, $X,\phi \in \mathcal{R}$. The equation for $\chi$ can then be decomposed into two equations with real coefficients yielding (here and throughout the manuscript we neglect the exciton damping $\gamma_x$ assuming it to be much smaller than all characteristic energies of the system):
\begin{align}
&{\delta}^{\prime}_X X+{g}^{\prime}X^3-(1-3\lambda X^2){P}_C\cos(\varphi-\varphi_0)-(X-4\lambda X^3+3\lambda X^5)\cos\varphi_0=0, \label{stat01} \\
&\sin(\varphi-\varphi_0)=\frac{X}{{P}_C}\left[(1-\lambda X^2)\sin\varphi_0-{{P}_X\alpha_0}\right] \label{stat02},
\end{align}
where $\alpha_0=\sqrt{{\delta}_C^2+{\gamma}_C^2}$, $\{{g}^{\prime},{\delta}^{\prime}_X\}=\{{g},{\delta}_X\}\times\sqrt{{\delta}_C^2+{\gamma}_C^2}$.

Equations (4), (5) are the subject of our analysis below.

\section{Stationary solutions.}\label{sec2}
At first, we analyse the stationary solutions of Eqs.~\eqref{stat01},\eqref{stat02}.

 We are looking for the real solutions for $X \in [0;1]$ and $\varphi \in [0; 2\pi[$. We start from the simplest case of only the optical pumping, setting  ${P}_X=0$ in Eqs.~\eqref{stat01},\eqref{stat02}. Typical results for this limit are displayed in Fig.~\ref{fig_2}(a). In particular, Eq.~\eqref{stat02} has two real solutions for $\varphi$ for any value of $X$:
\begin{align}
\varphi=\varphi_0\pm \sin^{-1}\left[\frac{X}{{P}_C}(1-\lambda X^2)\sin\varphi_0\right]+\frac{\pi}{2}(1\mp 1)
\end{align}
In order to elucidate the multistability regime we here adopt the assumption of off-resonant pumping and high-Q cavity, assuming ${\gamma}_C\ll {\delta}_C$. We also assume zero detuning between the exciton and photon mode leading to $\delta_C=\delta_X=\delta$ Within this approximation, $\sin\phi_0\approx 0$, $\cos\phi_0\approx \mathrm{signum}(\delta)$, and $\cos (\phi-\phi_0)\approx \pm\cos\phi_0$. Eq.~\eqref{stat01} thus reduces to 
\begin{align}
3\lambda X^5-({g}\delta+4\lambda)X^3-({\delta}^2-1) X\pm(1-3\lambda X^2){P}_C=0.\label{stat11}
\end{align}

Let us note that in the absence of the losses, the threshold for the multistability appearance should be obtained at arbitrary small pumping intensity. Thus, it is instructive to analyze the limiting case ${P}_C=0$. Then,  in the non-saturable case non-trivial solutions of Eq.~\eqref{stat11} satisfy the condition
\begin{align}
g\delta X^2+(\delta^2-1)=0, \label{nonsat}
\end{align}
and for the case with $\lambda=1$ we get
\begin{align}
3X^4-({g}\delta+4)X^2-(\delta^2-1)=0. \label{stat2}
\end{align}
For the nonsaturable case of Eq.~\eqref{nonsat} we can readily obtain the condition for the bistability existence, since a positive root exists only if $\delta \in ]-\infty,-1[\cup ]0;1[$. Thus in the non-saturable system the bistability can arise only if the resonant pump is either above the upper polariton frequency or between the lower polariton frequency and the exciton frequency.

For the saturable case of Eq.~\eqref{stat2}, there are three distinct cases: the equation  has no roots in the region $]0;1[$ and thus there is no multistable behaviour in the system. If it has one root, then  system will be bistable. Finally, if it has two roots, the tri-stability may be observed. In stark contrast to the non-saturable case, the tri-stability is observed in the region $\delta\in ]-1;0[$ i.e. when the resonant pump is between the exciton and upper polariton frequency, where no bistable behaviour is observed for the case of $\lambda=0$. We thus observe that the multistability behaviour originates merely due to the saturation nonlinearity.

In order to support the approximate analytical results we have performed the numerical calculation of the stationary solutions of the system. For the numerical calculations we use the following set of parameters: the Rabi-splitting $\Omega_R$ is set to 10 meV, usual for the GaAs based structures, the photon lifetime is set to $2$ ps which can be easily achieved in the state of the art high-quality microcavities and corresponds to $\gamma_C/\Omega_R=0.03$; the binding energy $E_B\approx 4$ meV which is usual for the GaAs structures and the Bohr radius $a_B=10$ nm. The optical pumping intensity is bound by approximately $1$ kW/cm$^2$ which corresponds to the dimensionless ${P}\approx 1$ and the electrical pumping current is bound by approximately $100 \mu$A which corresponds to ${P}_X\approx 1$. These values of optical pump intensities and currents are easily achievable in the state of the art microcavity set ups~\cite{BS_exp_elec}. We take $\lambda=1$ and compare the results with the case $\lambda=0$. The results of the numerical calculations are shown in Figs.~\ref{fig_2}(a).

In Fig.~\ref{fig_2}(a) one can see that for the detunings lying in the region $]-\Omega_R;0[$ there is no bistable behaviour in the non-saturable case (green dotted line), but we observe the tri-stability for $\lambda=1$ (green solid line).

Despite the presence of the multistability regime in the phase diagram of the system, it is problematic to construct an experimental protocol exhibiting the multistable behaviour. this is connected to the large ${P}_C$ behaviour of the system. Namely, as can be seen from Eq.~\eqref{stat11} in this limit the amplitude of the exciton field is approaching $1/\sqrt{3}$. At this limit, exciton and photon subsystems become decoupled and increasing the photon pumping will not affect the exciton subsystem. This asymptote is shown with a dashed line in Fig.~\ref{fig_2}(a). Thus it is not possible to switch between different stable exciton states by adiabatic change of the optical pump.

We now consider the case of simultaneous electrical and optical pump acting upon the system. One of the  effects observed in this case, is the shift of the $|{\chi}|$ vs ${P}_C$ dependence to the larger values of ${P}_C$ as ${P}_X$ is increased. To illustrate that we can  expand Eqs.~\eqref{stat01},\eqref{stat02} for small and large ${P}_C$ respectively. We again assume limit of low radiative losses $\gamma_C\ll \delta_C$, i.e. $\sin\phi_0\approx 0$. In this case, Eq.~\eqref{stat02} yields $\sin\phi=-X{P}_X|{\delta}_C|/{P}_C$. The requirement for the real phase $\phi$ leads to the condition $X<{P}_C/|({\delta}_C|{P}_X)$. Thus, as electrical pump increases, the exciton amplitude will decrease slower with the optical pump intensity. This also can be observed in Fig.~\ref{fig_2}(c) where the dependence of the exciton amplitude vs electric pump for different values of ${P}_C$ is shown. We can see that at the low branch of the stationary solutions, lying below the asymptote $|{\chi}|=1/\sqrt{3}$  the exciton amplitude decays with increasing the pump intensity. In order to better illustrate the exciton amplitude dependence on the electrical and optical pump, the three dimensional plot of $|\chi|$ vs $P_C$ and $P_X$ is shown in Fig.~\ref{fig_2}(d).

\section{Stability Analysis}\label{sec3}
\begin{figure*}[t]
\centerline{\includegraphics[width =
1.0\textwidth]{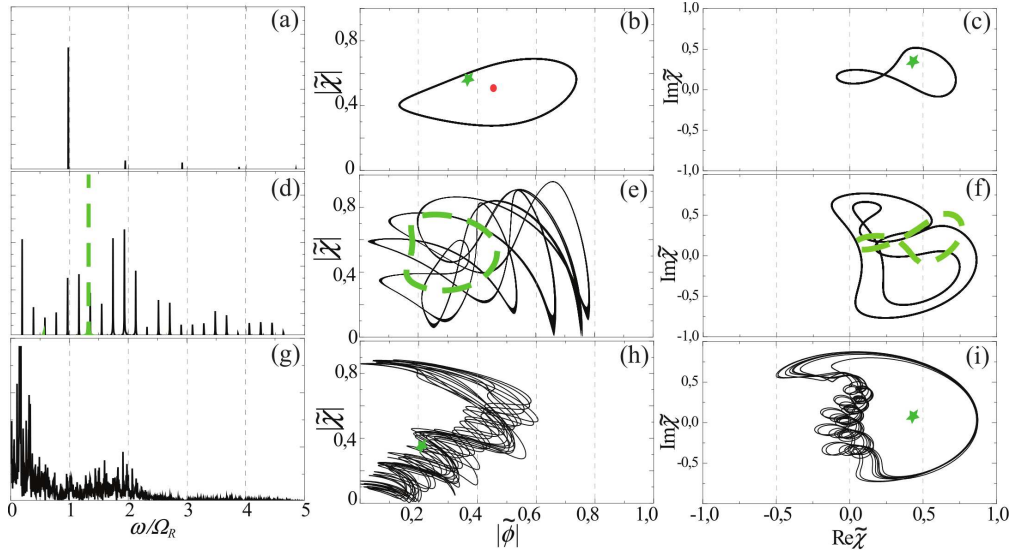}} \caption{Spectrum (left column  and  phase space $(|{\phi}(t)|,|{\chi}(t)|)$ (middle column) and $(\mathrm{Re}{\phi}(t),\mathrm{Im}{\chi}(t))$ (right column) trajectories  of the system at $\delta_C=-9$ meV and ${P}_C=0.5,\quad {P}_X=0.55$ (a,b,c), ${P}_C=0.5,\quad {P}_X=0.27$ (d,e,f), ${P}_C=0.2,\quad {P}_X=0.09$ (g,h,i). In all figures solid black lines correspond to saturable case $\lambda=1$, and dashed green to $\lambda=0$. In figure (b) red point corresponds to the stationary solution for the saturable case.}   \label{fig_3}
\end{figure*}
\begin{figure}[!h]
\centerline{\includegraphics[width =
0.4\columnwidth]{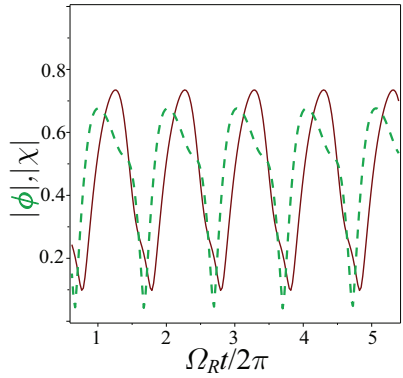}} \caption{ Time dependance the of the exciton and photon field at the same parameters as in Fig.~\ref{fig_3}(a) for $\lambda=1$.}  \label{fig_4}
\end{figure}

To analyze the stability of the obtained stationary solutions we have computed the Jacobian of the system at the stationary points~\cite{DynamicalAnalysis}. The appearance of the positive real parts of at least one eigenvalue of the Jacobian indicates the unstable behaviour. The stable stationary points are shown in bold in Figs.~\ref{fig_2}(a-c). We have focused at the regime of the detuning $\delta$ lying in the region $[-\Omega;0]$ corresponding to the tri-stable behaviour in the saturable case. We have identified the three types of the non-stationary dynamics of the system which are shown in Figs.\ref{fig_3}.
The first type is the periodic oscillations of the excitonic  and photonic components of the condensate. This behaviour is corresponding to the limiting cycle~\cite{DynamicalAnalysis} and is shown at Figs.~\ref{fig_3}(a,b,c). Relevant time dependent behaviour of photonic $|\phi|$ and excitonic $|\chi|$ amplitudes is shown at Fig.~\ref{fig_4}. Physically, this regime occurs due to the balance between dissipation and pumping in the presence of the nonlinearity and can be recognized as permanent Rabi oscillations in the condensate~\cite{Permanent1,Permanent2}. For the non-saturable case $\lambda=0$ with the same parameters, the system is stable.

The transition from the stationary solutions to the limiting cycle occurs as we decrease the electrical pump ${P}_X$ and as shown at Fig.~\ref{fig_2}(c). The spectrum of the oscillations is shown at Fig.~\ref{fig_3}(a). It can be seen that the oscillations occur primarily at the frequency of $\tilde{\Omega }\approx 0.95\Omega_R$. However, the small peaks corresponding to the multiples of ${\Omega}$ are also present. The trajectories in the phase space projections $(|\phi|,|\chi|)$ and $(\mathrm{Re}\chi,\mathrm{Im}\chi)$ in this case are  closed contours as it is illustrated in Figs.~\ref{fig_3}(b,c). We also observed that the closed contour contains the stationary point of the dynamical system which signifies the stability of the limiting cycle.

Further decreasing of  ${P}_X$ leads to the period multiplication in the system which is shown in Figs.~\ref{fig_3}(d,e,f). The spectrum that is shown in Fig.~\ref{fig_3}(d) exhibits the peaks at the frequencies $5\tilde{\Omega}/5$. We note that transition to 5th order multiplication of the period of the structure is sharp i.e. the system abruptly switches to the period multiplication by a factor of 5 rather than by consecutive period doubling. In this case the trajectories shown in Fig.~\ref{fig_3}(e,f) are still  closed contours while with far more complex topology. For the case of $\lambda=0$ with the same parameters, the transition to the permanent Rabi oscillations, i.e. limiting cycle is observed.

As the electrical pump is decreased further the system undergoes the transition to chaotic behaviour through a series of consecutive period multiplication. The spectrum in this case, shown in Fig.~\ref{fig_3}(g) becomes effectively a continuous one and the trajectories shown in Fig.~\ref{fig_3}(h,i) become unclosed. For the non-saturable case $\lambda=0$ at these parameters the system is stable again.

We have also observed a threshold of the optical pump intensity of $0.06$, below which the solutions are stable regardless of the electrical pump intensity. For the non-saturable case the threshold corresponds to the $0.11$.

\section{Conclusions}\label{sec4}

We analyzed the dynamics of the excitonic and photonic fields in a microcavity under simultaneous optical and electric pump accounting for the exciton-photon coupling saturation provided by Pauli exclusion principle. We have shown that in the limit of the vanishing electrical pump the system may exhibit the multistable behaviour at certain values of the detunings between the frequencies of the photon mode and the pump. Moreover, we observed the transition from the stable solutions to the chaotic behavior through the cascade of period multiplication as electrical pump in the system is tuned. The discovered regimes can be useful for further development of applications of polariton lasers.

\section{Acknowledgements}\label{sec5}

A.P.A. acknowledges useful discussions with Yuri Rubo. This work is supported under the project  No  RFMEFI58715X0020 of  the Federal Targeted Programme  “Research and Development in Priority Areas of Development of the Russian Scientific and Technological Complex for 2014-2020”  of  The  Ministry of Education and Science of Russia.. I.A.S. acknowledges support from Singapore Ministry of Education under AcRF Tier 2 grant MOE2015-T2-1-055.


\end{document}